\documentclass[twocolumn,floatfix,preprintnumbers,nofootinbib,superscriptaddress]{revtex4}

\usepackage{ulem}
\usepackage{bm}
\usepackage{times}
\usepackage{amssymb,amsbsy,amsmath,amsfonts}
\usepackage{graphicx}
\usepackage{float}
\usepackage{color}
\usepackage{morefloats}
\usepackage{rotating}
\usepackage{srcltx}
\usepackage{slashed}
\usepackage{subfigure}
\usepackage{multirow}
\usepackage{verbatim}
\usepackage{hyperref}
\usepackage{tabularx}

\usepackage[outdir=./]{epstopdf}

\usepackage{graphicx}
\usepackage{amsmath}
\usepackage{amsfonts}
\usepackage{amssymb}
\usepackage{color}
\usepackage{multirow}
\usepackage{mathrsfs}
\usepackage{gensymb}

\usepackage{subfigure}

\newcommand\be{\begin{eqnarray}}
\newcommand\ee{\end{eqnarray}}

\begin{document}

\title{Theoretical study on the contributions of $\omega$ meson to the $X(3872) \to J/\psi\pi^+\pi^-$ and $J/\psi\pi^+\pi^-\pi^0$ decays}

\author{Hao-Nan Wang}~\email{wanghaonan@m.scnu.edu.cn}
\affiliation{Guangdong Provincial Key Laboratory of Nuclear Science, Institute of Quantum Matter,
South China Normal University, Guangzhou 510006, China}
\affiliation{Institute of Modern Physics, Chinese Academy of Sciences, Lanzhou 730000, China}

\author{Qian Wang}~\email{qianwang@m.scnu.edu.cn}
\affiliation{Guangdong Provincial Key Laboratory of Nuclear Science, Institute of Quantum Matter,
South China Normal University, Guangzhou 510006, China}
\affiliation{Guangdong-Hong Kong Joint Laboratory of Quantum Matter, Southern Nuclear Science Computing Center,
South China Normal University, Guangzhou 510006, China}
\affiliation{Theoretical Physics Center for Science Facilities, Institute of High Energy Physics, Chinese Academy of Sciences,
Beijing 100049, China}

\author{Ju-Jun Xie}~\email{xiejujun@impcas.ac.cn}
\affiliation{Institute of Modern Physics, Chinese Academy of Sciences, Lanzhou 730000, China}
\affiliation{School of Nuclear Sciences and Technology, University of Chinese Academy of Sciences, Beijing 101408, China}
\affiliation{School of Physics and Microelectronics, Zhengzhou University, Zhengzhou, Henan 450001, China}

\begin{abstract}

With the newly measurements of $X(3872) \to J/\psi\pi^+\pi^-$ from LHCb Collaboration, in addition to the dominant contribution from the $\rho^0$ meson, we perform a theoretical study on the contribution of $\omega$ meson in the $X(3872) \to J/\psi\pi^+\pi^-$ and $X(3872)\to J/\psi\pi^+\pi^-\pi^0$ processes. It is found that the recent experimental measurements on the $\pi^+ \pi^-$ invariant mass distributions can be well reproduced, and the ratio of the couplings between $X(3872) \to J/\psi\rho^0$ and $X(3872) \to J/\psi\omega$ is also evaluated. Within the parameters extracted from the $\pi^+ \pi^-$ invariant mass distributions of the $X(3872) \to J/\psi\pi^+\pi^-$  process, the branching fractions of the $X(3872) \to J/\psi \omega$ channel relative to that of the $X(3872) \to J/\psi \pi^+\pi^-\pi^0$ channel and the $\pi^+ \pi^- \pi^0$ invariant mass distributions of the $X(3872) \to J/\psi \pi^+\pi^-\pi^0$ decay are calculated, which gives a hint for the further high-statistic experiment. 

\end{abstract}

\maketitle

\section{Introduction} \label{sec:Introduction}

In 2003, the $\chi_{c1}(3872)$ (also known as $X(3872)$), as an exotic candidate, was discovered by the Belle experiment in the $J/\psi \pi^+\pi^-$ channel~\cite{Belle:2003nnu}. An updated analysis was done in 2011~\cite{Belle:2011vlx}. Ten years after its observation, its quantum number has been well determined to be $I^G(J^{PC}) = 0^+(1^{++})$~\cite{LHCb:2013kgk}. Nowadays the $X(3872)$ is well established. The "OUR AVERAGE" value in the 2020 version and 2021 updated of the Review of Particle Physics (RPP)~\cite{ParticleDataGroup:2020ssz} for the mass of $X(3872)$ is $3871.65\pm0.06$ MeV, which is near the $D^0\bar{D}^{*0}$ mass threshold. As its mass is much lower than the one of $\chi_{c1}(2P)$ predicted by the $c\bar{c}$ quark model~\cite{Godfrey:1985xj}, it cannot be accepted by normal charmonium picture. Furthermore, its width is extremely narrow compared with other hadrons that have similar energy. In 2020, its Breit-Wigner width was measured and it is $1.39 \pm 0.24 \pm 0.10$ MeV~\cite{LHCb:2020xds} or $0.96^{+0.19}_{-0.18} \pm 0.21$ MeV~\cite{LHCb:2020fvo}, depending on the assumed lineshape. 

Although the $X(3872)$ have been intensively studied in various pictures,
for instance the $D\bar{D}^*+c.c.$ molecular picture, the compact tetraquark picture,
the normal charmonium with a mixture of the $D\bar{D}^*+c.c.$ molecule and so on. 
 The nature of the $X(3872)$ state is still puzzling (more details can be seen in these review articles~\cite{Lebed:2016hpi,Chen:2016qju,Esposito:2016noz,Olsen:2017bmm,Guo:2017jvc,Ali:2017jda,Brambilla:2019esw,Liu:2019zoy,Guo:2019twa,Dong:2021bvy,Chen:2022asf}). Because its mass is very close to the 
 $D^0\bar{D}^{*0}$ mass threshold, it could have a large $D^0\bar{D}^{*0}$ molecular component~\cite{Guo:2019qcn,Zhang:2020mpi}, which leads a large isospin breaking effect~\cite{Gamermann:2009fv,Li:2012cs,Meng:2021kmi,Tornqvist:2004qy,Takeuchi:2011uh,Terasaki:2009in,Takeuchi:2014rsa}. This isospin breaking effect has been found by Belle Collaboration~\cite{Belle:2011vlx}
in the ratio of three- and two-pion branching fractions
\be 
    R^{\mathcal{B}_{3\pi}/\mathcal{B}_{2\pi}} = \frac{\mathcal{B}[X(3872)\rightarrow J/\psi\pi^+\pi^-\pi^0]}{\mathcal{B}[X(3872)\rightarrow J/\psi\pi^+\pi^-]} = 0.8\pm 0.3.
    \label{formula:br}
\ee 
With $\mathcal{B}_{2\pi}=(3.8 \pm 1.2)\%$ quoted in RPP~\cite{ParticleDataGroup:2020ssz}, one can easily obtain $\mathcal{B}_{3\pi}=(3.0 \pm 1.5)\%$. If we assumed that the two-pion final state is dominated by the $\rho^0$ meson, while the three-pion is dominated by the $\omega$ meson, it is found that $X(3872)$ has large isospin violation effects in these two above decays, since they have the same rate within one sigma uncertainty.

The above two- and three-pion transition modes are our focus
due to the recent new measurements from LHCb collaboration~\cite{LHCb:2022bly}. 
As the $X(3872)$ decays to $J/\psi \pi^+\pi^-$ with $\pi^+\pi^-$ from $\rho^0$ meson~\cite{Belle:2003nnu,CDF:2005cfq}, the $X(3872) \to J/\psi \pi^+\pi^-$ is isospin breaking process due to the isospin singlet property to the $X(3872)$. The isospin conserving decay $X(3872) \to J\psi \omega \to J/\psi \pi^+\pi^- \pi^0$ is important to understand the internal structure of $X(3872)$, though it has small phase space. The $X(3872) \to J\psi \omega$ decay has been observed with a significance of more than $5\sigma$ by the BESIII collaboration~\cite{BESIII:2019qvy}. Previously, the Belle and BABAR collaborations also found evidence for the $X(3872) \to J/\psi \omega$ decay~\cite{Belle:2005lfc,BaBar:2010wfc}, but their measurements are with large uncertainties. Recently, sizeable $\omega$ contribution to $X(3872) \to J/\psi \pi^+\pi^-$ decay are observed by the LHCb collaboration~\cite{LHCb:2022bly}. These new measurements can be used to study the effects of the isospin-violating $\rho^0$ and $\omega$ mixing. Indeed, in Ref.~\cite{Hanhart:2011tn} this effect was taken into account in the analysis of the $J/\psi \pi^+\pi^-$ and $J/\psi \pi^+\pi^-\pi^0$ invariant mass distributions in the decays $X(3872) \to J/\psi \rho^0$ and $X(3872) \to J/\psi \omega$, where they focused on the determination of the quantum numbers of $X(3872)$. Besides, based on the $D\bar{D}^*$ molecular picture of $X(3872)$, the isospin breaking effects of the $X(3872) \to J/\psi \rho^0$ and $X(3872) \to J/\psi \omega$ decays were investigated in Refs.~\cite{Gamermann:2009fv,Li:2012cs,Meng:2021kmi,Suzuki:2005ha,Liu:2006df,Gamermann:2009uq,Coito:2010if,Albaladejo:2015dsa,Wu:2021udi}.

In this work, with the new measurements of the LHCb collaboration~\cite{LHCb:2022bly}, and following the work of Ref.~\cite{Hanhart:2011tn}, we study the invariant mass distributions of the $\pi^+\pi^-$ and $\pi^+\pi^-\pi^0$ final states in the $X(3872) \to J/\psi \pi^+\pi^-$ and $X(3872) \to J/\psi\pi^+\pi^-\pi^0$ decays, respectively, where we will focus on the role played by the $\omega$ meson to the $X(3872) \to J/\psi \pi^+\pi^-$ decay and the ratio of the effective couplings of $X(3872)$ to $J/\psi \rho^0$ and $J/\psi \omega$.

The paper is organized as follows. In Sec. II, we present the theoretical formalism of the $X(3872) \to J/\psi \pi^+\pi^-$ and $X(3872) \to J/\psi\pi^+\pi^-\pi^0$ decays, and in Sec. III, we show our numerical results and discussions, followed by a short summary in Sec. IV.


\section{formalism} \label{sec:Formalism}

The effective Lagrangian method is an useful tool in describing the various processes around the resonance region. The model used in the present work can give a reasonable description of the experimental data for the $X(3872) \to J/\psi \pi^+\pi^-$ decay, and our calculation offers some important clues for the mechanisms of the decays of $X(3872) \to J/\psi \pi^+\pi^-$ and $J/\psi \pi^+\pi^-\pi^0$. In this section, we introduce the theoretical formalism and ingredients to study the $X(3872) \to J/\psi \pi^+\pi^-$ and $X(3872) \to J/\psi \pi^+\pi^-\pi^0$ decays by using the effective Lagrangian method. 

\subsection{Feynman diagrams and effective interaction Lagrangian densities}

\begin{figure}[htbp]
\centering
\includegraphics[scale=0.7]{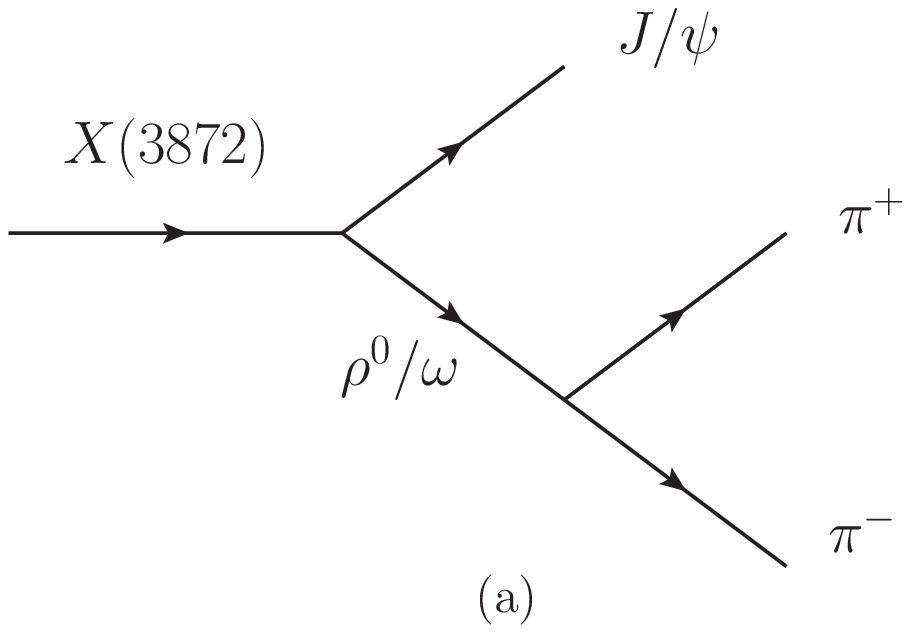}
\includegraphics[scale=0.7]{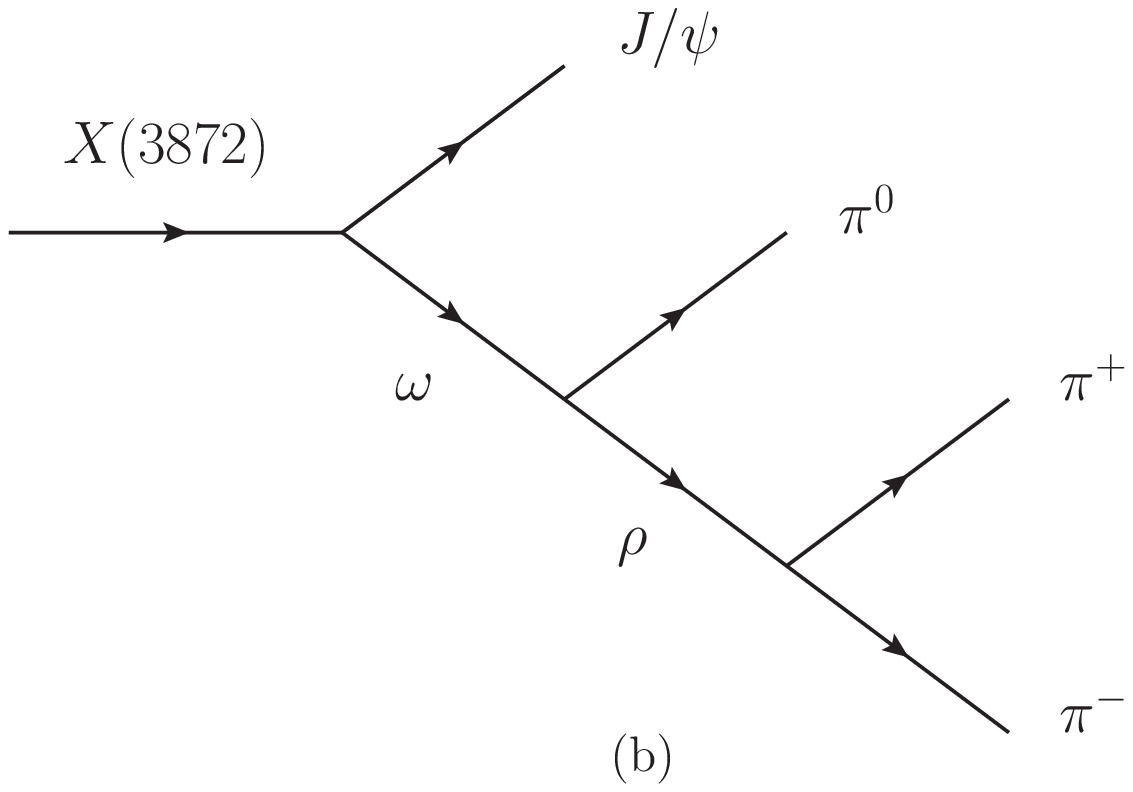}
\caption{Feynman diagrams for the $X(3872) \to J/\psi \pi^+\pi^-$ decay with $\rho^0$ and $\omega$ contribution (a) and $X(3872) \to J/\psi \pi^0 \pi^+\pi^-$ decay with $\omega$ contribution (b).} \label{fig:singlevector}
\end{figure}

Following previous analyses of Ref.~\cite{Hanhart:2011tn}, we assume that the $X(3872) \to J/\psi \pi^+\pi^-$ decay is mediated by the $\rho^0$ and $\omega$ meson, while the $X(3872) \to J/\psi \pi^+\pi^- \pi^0$ decay is mediated by the $\omega$ meson. The corresponding basic tree-level diagrams are shown in Fig.~\ref{fig:singlevector}. For the $\omega \to \pi^+\pi^-\pi^0$ decay, we take the $\rho$ meson as an intermediate state. The $\omega$ meson firstly couples to $\pi \rho$ and then the $\rho$ meson decays into $\pi\pi$ in the final state.~\footnote{In the calculation, we consider only the process of $\omega \to \pi^0\rho^0 \to \pi^0 \pi^+\pi^-$, and finally multiply by an isospin factor three to the total decay. This treatment will not change the three pion lineshape.} On the other hand, we also consider the contribution of $\omega$ meson to the $X(3872) \to J/\psi \pi^+\pi^-$ decay with $\omega$ decaying into $\pi^+\pi^-$. Note that, in Ref.~\cite{Hanhart:2011tn}, the $\omega \to \rho^0$ mixing was taken into account for the contribution of $\omega$ meson to the $\pi^+\pi^-$ production, where the transition amplitude is described by a real parameter. Here, we consider an effective coupling for the $\omega \to \pi^+\pi^-$ decay. 

To obtain the decay amplitudes of the processes shown in Fig.~\ref{fig:singlevector}, we need the effective interactions for these interaction vertexes, which can be described by the effective Lagrangian densities as used in Refs.~\cite{Lucio-Martinez:2000now,Xie:2008ts,Wang:2020duv,Janssen:1994uf}:
\begin{eqnarray}
    \mathcal{L}_{XJ/\psi V} &=& g_{X}\varepsilon^{\mu\nu\rho\sigma}\partial_{\mu}X_{\nu}J/\psi_{\rho} V_{\sigma}, \label{eq:XJpsiVlagrangian} \\
    \mathcal{L}_{\omega\rho\pi} &=& g_{\omega\rho\pi}\varepsilon^{\alpha\beta\gamma\tau}\partial_{\alpha}\omega_{\gamma}\partial_{\beta}\rho_{\tau}\phi_\pi, \label{eq:omegarhopionlagrangian}\\
 \mathcal{L}_{V\pi\pi} &=& g_{V\pi\pi} V_{\mu}\partial^{\mu}(\phi_{\pi^+}-\phi_{\pi^-}),
    \label{eq:Vpipilagrangian}
\end{eqnarray}
where $X$ and $V$ represent $X(3872)$ and $\rho^0/\omega$ meson, respectively. While $g_X$, $g_{\omega\rho\pi}$, $g_{\omega\pi\pi}$, and $g_{\rho\pi\pi}$ are the coupling constants of the corresponding vertexes. In this work, we will use $g_{X_1}$ and $g_{X_2}$ for the effective coupling constants of $X(3872)$ to $J/\psi \rho^0$ and $J/\psi \omega$, respectively. Furthermore, we will take $g_X$, $g_{\omega\rho\pi}$, $g_{\rho \pi\pi}$, and $g_{\omega\pi\pi}$ as real and their values will be discussed below. 

\subsection{Invariant decay amplitudes}

With the effective interaction Lagrangian densities
given above, the invariant decay amplitudes for these diagrams shown in Fig.~\ref{fig:singlevector} can be written as
\begin{eqnarray}
    \mathcal{M}_a &=& {\cal M}_{\rho} + {\cal M}_{\omega} \nonumber \\
    &=& -i \left ( \frac{g_{X_1}g_{\rho\pi\pi}}{D_{\rho}(q^2)} F_{\rho} (q^2) + \frac{g_{X_2}g_{\omega\pi\pi}}{D_\omega(q^2)} F_{\omega} (q^2) \right ) \times  \nonumber \\
    && \varepsilon^{\mu\nu\rho\sigma}p_{\mu}\epsilon_{\nu}(p)\epsilon_{\rho}^{\ast}(p_4)(p_1-p_2)_{\sigma}, \label{eq:amplitudeomega2pi} \\
    \mathcal{M}_b 
    &=& \frac{g_{X_2}g_{\omega\rho\pi}g_{\rho\pi\pi}}{D_{\omega}(q^2_1) D_{\rho}(q^2_2)}F_{\omega} (q_1^2)F_{\rho} (q_2^2) \varepsilon^{\mu\nu\rho\sigma}p_{\mu}\epsilon_{\nu}(p)\epsilon_{\rho}^{\ast}(p_4)\notag\\
    &&\times \left( -g_{\sigma\gamma}+\frac{q_{1\sigma}q_{1\gamma}}{m_{\omega}^2} \right) \varepsilon^{\alpha\beta\gamma\tau}  q_{1\alpha}q_{2\beta}(p_2-p_1)_{\tau},    
\end{eqnarray}
where $p$, $p_1$, $p_2$, $p_3$, $p_4$ are the four-momenta of $X(3872)$, $\pi^-$, $\pi^+$, $\pi^0$, $J/\psi$, while $q_1$ and $q_2$ represent the four-momenta for the intermediate $\omega$ and $\rho^0$ mesons. $D_{\rho}(q^2)$ and $D_{\omega}(q^2_1)$ are the denominators of the propagators for the $\rho^0$ and $\omega$ meson, which are
\begin{eqnarray}
D_{\rho}(q^2) & =& q^2-m_{\rho}^2+im_{\rho}\Gamma_{\rho} (q^2),  \label{Eq:rhowidth} \\
D_{\omega}(q^2_1) &=& q^2_1 -m^2_{\omega} + i m_{\omega} \Gamma_{\omega}.
\end{eqnarray}
Since the major decay channel of $\omega$ meson is the $\pi^+\pi^-\pi^0$ and its width is narrow, we take $m_\omega = 782.65$ MeV and $\Gamma_\omega = 8.49$ MeV as quoted in the Review of Particle Physics~\cite{ParticleDataGroup:2020ssz}. For the width of $\rho$ meson, since it is large and the predominant decay mode is $\pi\pi$, we take that $\Gamma_{\rho}$ is energy dependent, which is given by~\cite{Hanhart:2010wh,Zhang:2017eui}
\be 
    \Gamma_\rho(q^2) = \Gamma_0 \frac{|\textbf{p}_{\pi}|^3}{|\textbf{p}_{\pi}^0|^3}\frac{m_{\rho}^2}{q^2},
\ee 
where 
\be
    |\textbf{p}_{\pi}| = \frac{\sqrt{q^2 - 4m_{\pi}^2}}{2},~~~|\textbf{p}^0_{\pi}| = \frac{\sqrt{m_\rho^2-4m_{\pi}^2}}{2}.
\ee 

In evaluating the decay amplitudes of $X(3872) \to J/\psi \pi^+\pi^-$, we include the form factors for $\rho^0$ and $\omega$ mesons since they are not point like particles~\cite{Liu:1995st}. In this work we adopt here the common scheme used in many previous works~\cite{Xie:2008ts,Xie:2014kja,Xie:2014zga}:
\be
F_{\rho/\omega} (q^2) = \frac{\Lambda_{\rho/\omega}^4}{\Lambda_{\rho/\omega}^4+(q^2-m_{\rho/\omega}^2)^2}, \label{eq:formfactor}
\ee
where we assume that $\Lambda_{\rho} = \Lambda_\omega = \Lambda$ and they are determined to fit with the recent LHCb measurements. Note that a Blatt-Weisskopf barrier factor was used for the $P$-wave decay of a vector to $\pi\pi$ in Ref.~\cite{LHCb:2022bly}.

Besides, the coupling constants, $g_{\rho\pi\pi}$, $g_{\omega\pi\pi}$, and $g_{\omega\rho\pi}$, are determined from the experimentally observed partial decay widths of $\rho \to \pi\pi$, $\omega \to \pi\pi$, and $\omega \to \rho \pi \to \pi\pi\pi$, respectively. With the effective interaction Lagrangians shown in Eq.~\eqref{eq:Vpipilagrangian}, these partial decay widths $\rho \to \pi \pi$ and $\omega \to \pi\pi$ can be easily calculated. The coupling constants are related to the partial decay widths as
\begin{eqnarray}
\Gamma_{\rho \to \pi\pi} &=& \frac{g^2_{\rho\pi\pi}}{6\pi} \frac{p_{\rho}^3}{m^2_{\rho}}, \\
\Gamma_{\omega \to \pi\pi} &=& \frac{g^2_{\omega\pi\pi}}{6\pi} \frac{p^3_\omega}{m^2_{\omega}},
\end{eqnarray}
where $p_{\rho}$ and $p_\omega$ are the three momenta of the $\pi$ meson in the $\rho$ or $\omega$ rest frame, respectively. With $\Gamma_{\rho \to \pi\pi} = 149.1$ MeV and $\Gamma_{\omega \to \pi\pi} = 0.133$ MeV as quoted in the Ref.~\cite{ParticleDataGroup:2020ssz}, we obtain $g_{\rho\pi\pi}=5.97$ and $g_{\omega \pi\pi} = 0.18$, respectively. Note that from the partial decay width, one can only obtain the absolute value of the coupling constant, but not the phase. In this work, we assume that $g_{\rho\pi\pi}$ and $g_{\omega\pi\pi}$ are real and positive. In fact, these values obtained here were used in Refs.~\cite{Janssen:1994uf,Janssen:1996kx,Meissner:1987ge} for other processes. 

In Ref.~\cite{Chen:2017jcw}, it was found that that $\omega$-$\rho^0$ mixing
plays the major role in the evaluating the partial decay width of $\omega \to \pi^+\pi^-$, and
its contribution is two orders of magnitude larger than that from the direct $\omega \pi\pi$ coupling. However, here we obtained the coupling constant $g_{\omega\pi\pi}$ with the experimental results of the $\omega \to \pi^+\pi^-$ decay. In other words, we have taken the effective $\omega\pi\pi$ coupling as a constant, and determined with the experimental partial decay width of $\omega \to \pi\pi$, rather than the mixing between $\rho^0-\omega$ with the explicit $\rho^0$ propagator~\cite{Chen:2017jcw}. 

In addition, the value of $g_{\omega\rho\pi}$ is determined with the partial decay width of $\omega \to \pi^+\pi^-\pi^0$, which reads
\begin{eqnarray}
\text{d}\Gamma_{\omega \to \pi^+\pi^-\pi^0} = \frac{1}{(2\pi)^3}\frac{1}{32m_{\omega}^3}|\mathcal{M}|^2\text{d}M_{\pi^+\pi^-}^2\text{d}M_{\pi^-\pi^0}^2,
\end{eqnarray}
where
\be 
    \mathcal{M} =  \frac{-ig_{\omega\rho\pi}g_{\rho\pi\pi}}{D_{\rho}(q^2_{\rho})}\varepsilon_{\mu\nu\rho\sigma}q^{\mu}_{\omega}q_{\rho} \epsilon^{\nu}(q_\omega)(p_{\pi^+}-p_{\pi^-})^{\sigma},
\ee 
where $q_\omega$, $q_\rho$, $p_{\pi^+}$, and $p_{\pi^-}$ stand for the four-momenta of $\omega$, $\pi^+$, and $\pi^-$, respectively. With $\Gamma_{\omega \to \pi^+\pi^-\pi^0} = 7.74$ MeV, we obtain $g_{\omega\rho\pi}=0.046$ ${\rm MeV}^{-1}$ for the case of $\Gamma_{\rho}$ energy dependent and $g_{\omega\rho\pi}=0.05$ ${\rm MeV}^{-1}$ for the case of $\Gamma_\rho$ as a constant. One see that the affect of the $\Gamma_{\rho}$ energy dependent is rather small and can be neglected.

\subsection{Invariant mass distributions}

With the formalism and ingredients given above, the calculations of the invariant mass distribution for the $X(3872) \to J/\psi \pi^+\pi^-$ and $X(3872) \to J/\psi \pi^+ \pi^- \pi^0$ decays are straightforward~\cite{ParticleDataGroup:2020ssz}. The invariant $\pi^+\pi^-$ mass distribution of the $X(3872) \to J/\psi \pi^+\pi^-$ decay is given by
\begin{eqnarray} 
    &&\frac{\text{d}\Gamma_{X{3872} \to J/\psi \pi^+ \pi^-}}{\text{d}M_{\pi^+\pi^-}} = \frac{1}{24(2\pi)^4M_X^2} \times \notag\\ 
    && \int\Sigma|\mathcal{M}_{2\pi}|^2|\bold{p}_1^{\ast}||\bold{p}_4|\text{d}\cos{\theta_1}\text{d}\phi_1, \label{formula:dgdm23}
    \end{eqnarray}
where $\bold{p}_1^{\ast}$ and ($\theta_1$, $\phi_1$) are the three-momentum and
decay angle of the outing $\pi^+$ (or $\pi^-$) in the
center-of-mass (c.m.) frame of the final $\pi^+ \pi^-$ system,
$\bold{p}_4$ is the three-momentum of the final $J/\psi$ meson in the
rest frame of $X(3872)$, and $M_{\pi^+ \pi^-}$ is the
invariant mass of the final $\pi^+ \pi^-$ system.

For the invariant $\pi^+\pi^-\pi^0$ mass distributions of the $X(3872) \to J/\psi \pi^+\pi^-\pi^0$ decay, it is given by,
    \begin{eqnarray}
     &&\frac{\text{d}\Gamma_{X(3872) \to J/\psi \pi^+\pi^-\pi^0}}{\text{d}M_{\pi^+\pi^-\pi^0}} = \frac{1}{16(2\pi)^7M_X^2} \int\Sigma|\mathcal{M}_{3\pi}|^2  \times \notag \\  
     && |\bold{p}_1^{\ast}||\bold{p}_3'||\bold{p}_4| \text{d}M_{\pi^+\pi^-}\text{d}\cos{\theta_1}\text{d}\phi_1\text{d}\cos{\theta_2}\text{d}\phi_2,  \label{formula:dgdm123} 
\end{eqnarray}
with $M_{\pi^+\pi^-\pi^0}$ the invariant mass of $\pi^+\pi^-\pi^0$ system. The definitions of these variables in the phase space integration are given in Appendix A.

Besides, in Eqs.~\eqref{formula:dgdm23} and \eqref{formula:dgdm123}, we take
\begin{eqnarray}
{\cal M}_{2\pi} &=& {\cal M}_\rho + e^{i \varphi}{\cal M}_\omega, \\
{\cal M}_{3\pi} &=& {\cal M}_b.
\end{eqnarray}
Note that we have included a free parameter $\varphi$ which stands for the relative phase between $\omega$ and $\rho^0$ terms for the $X(3872) \to J/\psi \pi^+\pi^-$ decay. On the other hand, more details for the integration of the multi-body phase space can be found in Refs.~\cite{Xie:2015zga,Xie:2018gbi,Jing:2020tth}.


\section{Numerical results and discussions} \label{sec:Results}

In this section, we present the numerical results for the invariant mass distribution of $\pi^+\pi^-$ of the $X(3872) \to J/\psi \pi^+\pi^-$ decay. To compare the theoretical invariant mass distributions with the experimental measurements, we introduce an extra global normalization factor $C$, which will be fitted to the experimental data. In the calculation, the masses, widths and spin-parities of the involved particles are listed in Table~\ref{tab:particleparameters}.

\begin{table}[htbp]
\caption{Masses, widths and spin-parities of the involved particles in this work.}	\label{tab:particleparameters}
	\begin{tabular}{cccc}
		\hline\hline  
		Particle & Mass (MeV) & Width (MeV) & Spin-parity ($J^P$) \\ \hline
		$X(3872)$ & 3871.69 & 1.19$\pm$0.21 & $1^+$ \\
		$J/\psi$  & 3096.9  & ---           & $1^-$ \\
		$\rho^0$  & 775.26  & 149.1$\pm$0.8 & $1^-$ \\
		$\omega$  & 782.66  & 8.68$\pm$0.13 & $1^-$ \\
		$\pi^+/\pi^-$   &139.57   & ---           & $0^-$ \\
		$\pi^0$   &134.97   & ---           & $0^-$ \\
		\hline\hline
	\end{tabular}
\end{table}

We perform four parameters ($R_X \equiv g_{X_1}/g_{X_2}$, $Cg_{X_2}^2$, $\Lambda$, and $\varphi$) $\chi^2$ fits to the experimental data on the $\pi^+\pi^-$ invariant mass distributions. We will study two types of fit: one takes the total width of $\rho$ energy dependent, while the other one takes the total width of $\rho$ as a constant. The fitted parameters and the corresponding $\chi^2/\mathrm{d.o.f.}$ are shown in Table~\ref{table:2pifittedparameters}. We have checked that the results of the two fits are very similar, this indicates that the affects of the energy dependent of the $\rho$ total width $\Gamma_{\rho}$ is very small and can be neglected. It is worth to mention that the obtained ratio $R_X$ is very similar with these values $0.29 \pm 0.04$ and $0.26^{+0.08}_{-0.05}$ obtained in Ref.~\cite{LHCb:2022bly} and Ref.~\cite{Hanhart:2011tn}, respectively.

Although the two parameters $R_X$ and $Cg_{X_2}^2$ can be obtained
 from the fit directly, the physical couplings $g_{X_1}$ and $g_{X_2}$
 can only be extracted with the further inputs. With the value $R_X=0.25\pm 0.01$,
 the coupling $g_{X_2}=0.33\pm 0.06$ can be extracted from the
 the branching ratio $\mathcal{B}(X(3872) \to J/\psi \pi^+\pi^-)=(3.8\pm1.2)\%$~\cite{ParticleDataGroup:2020ssz}.
Consequently, the coupling $g_{X_1}=0.08\pm 0.02$ and the 
normalization factor $C=(7.48\pm 2.78)\times 10^{-5}$ can be obtained. The above values have been listed in Table~\ref{table:2pifittedparameters} as well as those for the constant $\Gamma_\rho$ case.

\begin{table}[htbp]
\centering
		\caption{The fitted and determined parameters in this work. The second and third columns are the fitted results of the $\Gamma_{\rho}(q^2)$ energy-dependent fit and fixed $\Gamma_0$ fit, respectively.}  
		\label{table:2pifittedparameters} 
		\begin{tabular}{ccc}   
			\hline\hline 
			 ~~Number~~ & ~~1~~ & ~~2~~ \\
			 \hline
			 $\rho$ meson width & $\Gamma_{\rho}(q^2)$ energy dependent & $\Gamma_0$ constant \\
$R_X$ & $0.25 \pm 0.01$ & $0.30 \pm 0.01$  \\
			 $Cg_{X_2}^2 ~ (\times 10^4)$ & $8.31\pm0.45$ & $7.10\pm0.48$\\
			 $\Lambda$ (MeV) & $612 \pm 18$ & $598 \pm 16$ \\
			 $\varphi~ (\degree)$ & $128.9 \pm 8.0$ & $134.5 \pm 7.7$ \\
			 $\chi^2/\mathrm{d.o.f.}$ & $0.6$ & $0.6$\\
			 \hline
			 $g_{X_1}$ & $0.08 \pm 0.02$ & $0.09 \pm 0.02$\\
			 $g_{X_2}$ & $0.33 \pm 0.06$ & $0.31 \pm 0.06$\\
			 $C~(\times 10^5)$ & $7.48 \pm 2.78$ & $7.48 \pm 2.81$ \\
			\hline\hline   
		\end{tabular}    
\end{table}

\begin{figure*}[htbp]
    \centering
    {\includegraphics[scale=0.7]{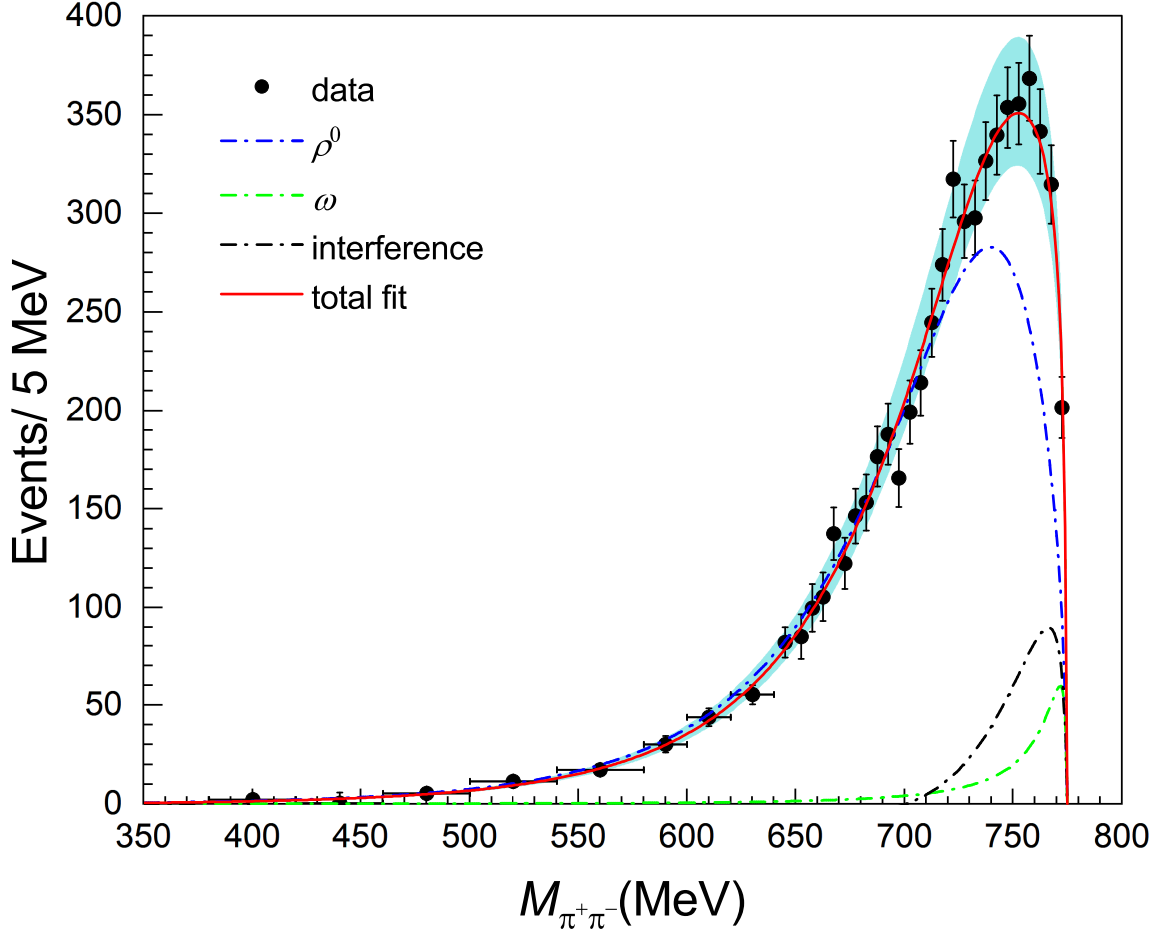}}  
    \caption{Invariant mass distribution of $\pi^+\pi^-$ for the $X(3872) \to J/\psi \pi^+\pi^-$ decay, compared with the experimental data taken from Ref.~\cite{LHCb:2022bly}. The blue-dash-dotted, green-dash-dotted curves and black-dash-dotted are the contributions from the $\rho^0$, $\omega$, and their interference, respectively. The red solid one is their total contribution. The band accounts for the corresponding $68\%$ confidence-level interval deduced from the distributions of the fitted parameters shown in Table~\ref{table:2pifittedparameters}}
    \label{fig:2pifitted}
\end{figure*}

With the central values of Table~\ref{table:2pifittedparameters} for the case of $\Gamma_{\rho}$ energy dependent, the $\pi^+\pi^-$ invariant mass distribution is shown by the red curve in Fig.~\ref{fig:2pifitted}. Note that the results for the case of $\Gamma_{\rho}$ as a constant are very similar with the ones obtained for the case of $\Gamma_{\rho}$ energy dependent. In Fig.~\ref{fig:2pifitted}, the red-solid curve stands for the total contributions from the $\rho^0$ and $\omega$ mesons, the blue-dash-dotted and green-dash-dotted curves correspond to the contribution from only the $\rho^0$ and $\omega$, respectively, while the black-dash-dotted stands for their interference. The band accounts for the corresponding $68\%$ confidence-level interval
deduced from the distributions of the fitted parameters shown in Table~\ref{table:2pifittedparameters}. One can see that the total numerical results can explain the experimental data quite well. Furthermore, the contribution of $\rho$ meson is predominant in the whole energy region consider for the $M_{\pi^+\pi^-}$, while the contribution of $\omega$ meson is crucial to the $\pi^+\pi^-$ invariant mass distributions at high energy of $M_{\pi^+\pi^-}$.

With the fitted couplings $g_{X_1}$ and $g_{X_2}$, one can easily obtain the $\omega$ contribution, without the $\rho^0$-$\omega$ interference terms, to the total $X(3872) \to J/\psi \pi^+\pi^-$ decay as
\begin{eqnarray}
\frac{\Gamma_{X(3872) \to J/\psi \omega \to J/\psi \pi^+\pi^-}}{\Gamma_{X(3872) \to J/\psi \pi^+\pi^-}} = (4.7 \pm 2.4)\%  \label{eq:omega1}
\end{eqnarray}
for the case of $\Gamma_{\rho}$ energy dependent and
\begin{eqnarray}
\frac{\Gamma_{X(3872) \to J/\psi \omega \to J/\psi \pi^+\pi^-}}{\Gamma_{X(3872) \to J/\psi \pi^+\pi^-}} = (4.0 \pm 2.1)\% \label{eq:omega2}
\end{eqnarray}
for the case of $\Gamma_\rho$ as a constant. These values agree with the value $(1.9 \pm 0.4 \pm 0.3)\%$, obtained by the LHCb analysis in Ref.~\cite{LHCb:2022bly} within one standard deviation. 

It is customary to apply the so-called narrow width approximation in the case where a particle decays into two particles and one of them with narrow width subsequently decays into other two particles in the final state~\cite{Cheng:2020iwk,Cheng:2020mna}. Since the width of $\omega$ meson is so narrow, we can extract the branching fraction of the quasi-two-body decay $X(3872) \to J/\psi \omega$ 
\begin{eqnarray}
&& {\cal B} (X(3872) \to J/\psi \omega) \nonumber \\
&=& \frac{{\cal B}(X(3872) \to J/\psi \omega \to J/\psi \pi^+\pi^-)}{{\cal B}(\omega \to \pi^+\pi^-)} \nonumber \\
&=& (11.5 \pm 7.5)\%,
\end{eqnarray}
within the narrow width approximation.
Furthermore, we extract the branching ratio fraction between the $J/\psi \omega$ and $J/\psi \pi^+\pi^-$ modes as defined in Ref.~\cite{BESIII:2019qvy}.
\begin{eqnarray}
\mathcal{R}\equiv \frac{{\cal B}(X(3872) \to J/\psi \omega )}{{\cal B}(X(3872) \to J/\psi \pi^+\pi^-)} = 3.0 \pm 2.2 \end{eqnarray}
for the $\Gamma_{\rho}$ energy-dependent case, and
\begin{eqnarray}
&& {\cal B} (X(3872) \to J/\psi \omega) = (9.9 \pm 6.3)\%, \\
&& \mathcal{R} = 2.6 \pm 1.9
\end{eqnarray}
for the constant $\Gamma_\rho$ case. The two values of $\mathcal{R}$ are in agreement with the experimental measurements $1.6^{+0.4}_{-0.3} \pm 0.2$ by BESIII collaboration~\cite{BESIII:2019qvy} within uncertainties.

Next we turn to the $X(3872) \to J/\psi \pi^+\pi^- \pi^0$ decay. With the values of $g_{X_1}$ and $g_{X_2}$, we obtain:
\begin{eqnarray}
{\cal B} [X(3872) \to J/\psi \pi^+\pi^-\pi^0] = (1.1 \pm 0.5)\%,
\end{eqnarray}
for the $\Gamma_{\rho}$ energy-dependent case and
\begin{eqnarray}
{\cal B} [X(3872) \to J/\psi \pi^+\pi^-\pi^0] = (1.4 \pm 0.6)\%,
\end{eqnarray}
for the constant $\Gamma_\rho$ case. Those values are in agreement with both the experimental measurements~\cite{ParticleDataGroup:2020ssz} and the theoretical calculations in Ref.~\cite{Aceti:2012cb}. 

\begin{figure}[htbp]
    \centering
    {\includegraphics[scale=0.36]{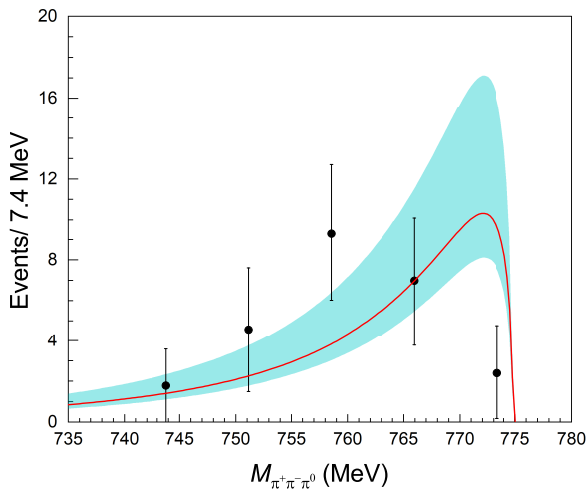}}  
    \caption{Invariant mass distribution of $\pi^+\pi^-\pi^0$ for the $X(3872) \to J/\psi \pi^+\pi^- \pi^0$ decay, compared with the experimental data taken from Ref.~\cite{BaBar:2010wfc}.}
    \label{fig:3pifit}
\end{figure}

Finally, with these model parameters determined by fitting to the $\pi^+\pi^-$ invariant mass distributions for the $X(3872) \to J/\psi \pi^+\pi^-$ decay, we calculate the $\pi^+\pi^-\pi^0$ invariant mass distributions for the $X(3872) \to J/\psi \pi^+\pi^- \pi^0$ decay. The numerical results are shown as the red curve in Fig.~\ref{fig:3pifit}. The error band of the theoretical calculations are obtained from the uncertainty of the parameter $g_{X_2}$ which stems from the uncertainties of both the two pion invariant mass distribution of the $J/\psi \pi^+\pi^-$ decay mode, i.e. the fitted parameter $R_X$, and the $X(3872)\to J/\psi \pi^+\pi^-$ branching ratio. To compare the theoretical invariant mass distributions with the experimental measurements, we have to introduce again an extra global normalization factor $C_{3\pi}$. In Fig.~\ref{fig:3pifit} the red-solid curve has been adjusted to the strength of the inverse-second data point of BABAR~\cite{BaBar:2010wfc} by taking $C_{3\pi} = 6.07 \times 10^4$ for the $\Gamma_{\rho}$ energy-dependent case. For the constant $\Gamma_\rho$ case, the value of $C_{3\pi}$ is $5.77\times 10^4$, and the line shape of the $\pi^+\pi^-\pi^0$ invariant mass distributions is almost the same. As shown in Fig.~\ref{fig:3pifit}, most of the experimental data locate in the theoretical one sigma 
region. This can be tested by future precise measurements for the $X(3872) \to J/\psi \pi^+\pi^- \pi^0$ channel. In addition, further more precise measurements of the
$X(3872) \to J/\psi \pi^+\pi^-$ channel can also help to reduce the uncertainty of the couplings $g_{X_1}$ and $g_{X_2}$. The $g_{X_2}$ can also constrain the data in $J/\psi \pi^+\pi^-\pi^0$ channel. 


\section{Summary} \label{sec:Conclusions}

In summary, we have performed a theoretical calculation for the processes of $X(3872) \to J/\psi\pi^+\pi^-$ and $J/\psi\pi^+\pi^-\pi^0$. For the $X(3872) \to J/\psi\pi^+\pi^-$ decay, in addition to the dominant contribution from the $\rho^0$ meson, the contribution of the intermediate $\omega$ meson with an effective $\omega \pi \pi$ coupling is also considered in our framework. It is found that the recent LHCb experimental measurements on the $\pi^+ \pi^-$ invariant mass distributions~\cite{LHCb:2022bly} can be well reproduced. Meanwhile, the ratio of the couplings between $X(3872) \to J/\psi\rho^0$ and $X(3872) \to J/\psi\omega$ is determined, which is consistent with the previous analysis in Refs.~\cite{Hanhart:2011tn,LHCb:2022bly}.

Furthermore, with the model parameters determined from the $\pi^+\pi^-$ invariant mass distribution of the $X(3872) \to J/\psi\pi^+\pi^-$ decay, the $X(3872) \to J/\psi\pi^+\pi^-\pi^0$ branching fraction and the corresponding $\pi^+ \pi^- \pi^0$ invariant mass distributions are extracted, which are also in agreement with the available experimental data with large errors. This kind of results could be tested by the future precise measurements.


\section*{Acknowledgments}

We would like to thank Prof. Gang Li for useful discussions. This work is partly supported by the National Natural Science Foundation of China under Grant Nos. ~12075288, ~11735003,~11961141012,~12035007, the Youth Innovation Promotion Association CAS, Guangdong Provincial funding with Grant No.~2019QN01X172, Science and Technology Program of Guangzhou No.~2019050001. Q.W. is also supported by the NSFC and the Deutsche Forschungsgemeinschaft (DFG, German Research Foundation) through the funds provided to the Sino-German Collaborative
Research Center TRR110 ``Symmetries and the Emergence of Structure in QCD"
(NSFC Grant No. 12070131001, DFG Project-ID 196253076-TRR 110).

\appendix

\section{Four-body phase space}

\begin{figure*}[htbp]
    \centering
    {\includegraphics[scale=0.7]{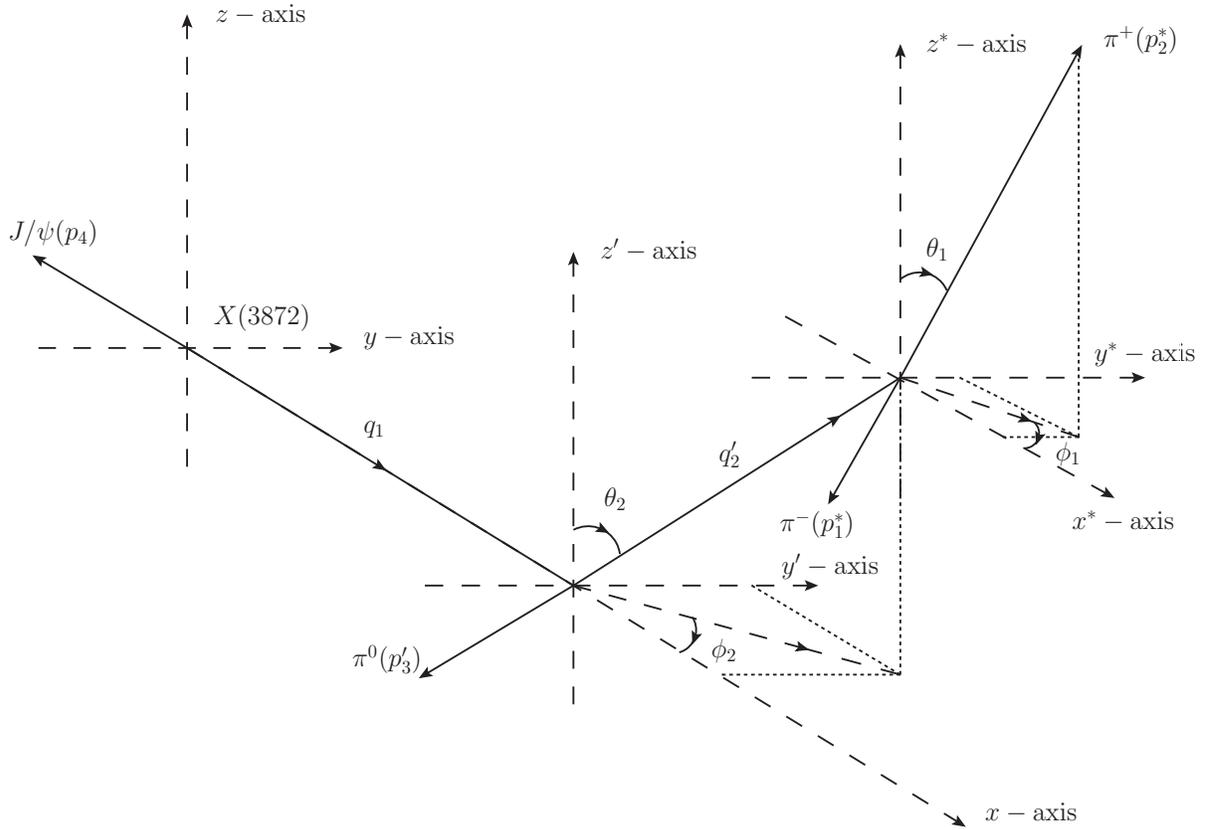}}  
    \caption{The definitions of these variables in the phase space integration of the $X(3872)\to J/\psi \omega \to J/\psi \pi^+\pi^-\pi^0$ decay.}
    \label{4bodyps}
\end{figure*}

In this appendix, we provide the definitions of those variables in the phase space integration of Eq.~\eqref{formula:dgdm123}, which are explicitly shown in Fig.~\ref{4bodyps}. The $\bold{p}_1^{\ast}$ and ($\theta_1$, $\phi_1$) are the three-momentum and decay angles of the outing $\pi^-$ in the $\pi^+ \pi^-$ center-of-mass (c.m.) frame. The $\bold{p}'_3$ and ($\theta_2$, $\phi_2$) are the three-momentum and decay angles of the outing $\pi^0$ in the $\pi^+ \pi^- \pi^0$ c.m. frame. The $\bold{p}_4$ is the three-momentum of the final $J/\psi$ meson in the $X(3872)$ rest frame.

\end{document}